\documentclass[sigplan,screen,10pt]{acmart}
\copyrightyear{2023}
\acmYear{2023}
\acmConference[PPoPP '23]{The 28th ACM SIGPLAN Annual Symposium on Principles and Practice of Parallel Programming}{February 25-March 1, 2023}{Montreal, QC, Canada}
\acmBooktitle{The 28th ACM SIGPLAN Annual Symposium on Principles and Practice of Parallel Programming (PPoPP '23), February 25-March 1, 2023, Montreal, QC, Canada}
\acmDOI{10.1145/3572848.3577434}
\acmISBN{979-8-4007-0015-6/23/02}
\startPage{1}

\setcopyright{rightsretained}
\usepackage{booktabs}   %% For formal tables:
\usepackage{subcaption} %% For complex figures with subfigures/subcaptions
\usepackage{natbib}     %% For Bib/Cites

\usepackage{comment}
\usepackage{alltt, listings, textcomp, color, verbatim}

\usepackage{xcolor}
\usepackage{ifthen}
\definecolor{frenchplum}{RGB}{129,20,83}
\definecolor{pastelpink}{RGB}{250,190,233}
\definecolor{pastelyellow}{RGB}{250,246,186}
\definecolor{pastelpurple}{RGB}{221,199,255}
\definecolor{pastelblue}{RGB}{183,239,255}
\definecolor{purple}{RGB}{159, 90, 253}

\newcommand{\final}{1}
\newcommand{\todo}   [1]{{{\color{frenchplum}\colorbox{frenchplum!30}{todo} #1}}}
\ifthenelse{\equal{\final}{1}}{\renewcommand{\todo}[1]{}}{}
\newcommand{\mosama}   [1]{{{\color{red}(mosama) #1}}}
\ifthenelse{\equal{\final}{1}}{\renewcommand{\mosama}[1]{}}{}
\newcommand{\jowens}   [1]{{{\color{blue}(jowens) #1}}}
\ifthenelse{\equal{\final}{1}}{\renewcommand{\jowens}[1]{}}{}
\newcommand{\serban}   [1]{{{\color{purple}(serban) #1}}}
\ifthenelse{\equal{\final}{1}}{\renewcommand{\serban}[1]{}}{}

\newcommand{\cpp}{\texorpdfstring{{C\nolinebreak[4]\hspace{-.05em}\raisebox{.10ex}{{++}}}}{C++}}

\usepackage[frozencache=true]{minted}
\usepackage[most]{tcolorbox}

\tcbset{textmarker/.style={%
        enhanced,
        parbox=false,boxrule=0mm,boxsep=0mm,arc=0mm,
        outer arc=0mm,left=4mm,right=3mm,top=7pt,bottom=7pt,
        toptitle=1mm,bottomtitle=1mm,oversize}}

\newtcolorbox{evaluation_box}{textmarker,
    borderline west={6pt}{0pt}{pastelyellow},
    colback=pastelyellow!10!white}

\newtcolorbox[auto counter]{sidebar_box}[2][]{textmarker,
    floatplacement=t,
    float,
    borderline west={6pt}{0pt}{pastelblue},
    colback=pastelblue!10!white,
    text width=\columnwidth,
    halign=justify,
    title=\textcolor{black}{\textbf{Sidebar~\thetcbcounter}~#2},
    title code={
      \path[fill=pastelblue!10!white] (title.south west) rectangle (title.north east);
      \path[draw=pastelblue,solid,line width=0.75mm]
      ([xshift=0mm]title.south west) -- ([xshift=0mm]title.south east);
      },
    nameref={#2},
    #1
}

\usepackage{float,ragged2e}
\floatstyle{boxed}
\begin{document}

%% Title information
% \title[Short Title]{Programming Model for GPU Load Balancing}
\title{A Programming Model for GPU Load Balancing}
                                        %% [Short Title] is optional;
                                        %% when present, will be used in
                                        %% header instead of Full Title.
% \titlenote{with title note}           %% \titlenote is optional;
                                        %% can be repeated if necessary;
                                        %% contents suppressed with 'anonymous'
% \subtitle{Subtitle}                   %% \subtitle is optional
% \subtitlenote{with subtitle note}     %% \subtitlenote is optional;
                                        %% can be repeated if necessary;
                                        %% contents suppressed with 'anonymous'

%% Author information
%% Contents and number of authors suppressed with 'anonymous'.
%% Each author should be introduced by \author, followed by
%% \authornote (optional), \orcid (optional), \affiliation, and
%% \email.
%% An author may have multiple affiliations and/or emails; repeat the
%% appropriate command.
%% Many elements are not rendered, but should be provided for metadata
%% extraction tools.

%% Author with single affiliation.
\author[M. Osama]{Muhammad Osama}
\email{mosama@ucdavis.edu}
\orcid{0000-0003-1616-6817}
\affiliation{%
  \institution{University of California, Davis}
  \streetaddress{1 Shields Ave}
  \city{Davis}
  \state{California}
  \country{USA}
  % \postcode{95616}
}

\author[S. D. Porumbescu]{Serban D. Porumbescu}
\email{sdporumbescu@ucdavis.edu}
\orcid{0000-0003-1523-9199}
\affiliation{%
  \institution{University of California, Davis}
  \streetaddress{1 Shields Ave}
  \city{Davis}
  \state{California}
  \country{USA}
  % \postcode{95616}
}

\author[J. D. Owens]{John D. Owens}
\email{jowens@ucdavis.edu}
\orcid{0000-0001-6582-8237}
\affiliation{%
  \institution{University of California, Davis}
  \streetaddress{1 Shields Ave}
  \city{Davis}
  \state{California}
  \country{USA}
  % \postcode{95616}
}

% conference papers do not typically use \thanks and this command
% is locked out in conference mode. If really needed, such as for
% the acknowledgment of grants, issue a \IEEEoverridecommandlockouts
% after \documentclass

\thanks{Distribution Statement ``A'' (Approved for Public Release, Distribution Unlimited).}

%% Abstract
%% Note: \begin{abstract}...\end{abstract} environment must come
%% before \maketitle command
\begin{abstract}
    % We propose to build an open-source GPU load-balancing framework for applications that exhibit irregular parallelism. The set of applications and algorithms we consider are fundamental to computing tasks ranging from sparse machine learning, large numerical simulations, and on through to graph analytics. The underlying data and data structures that drive these applications exhibit access patterns that naturally don't map well to the GPU's architecture that is designed with dense and regular access patterns in mind. Prior to the work we present and propose here, the only way to unleash the GPU's full power on these problems has been to workload balance through tightly coupled load-balancing techniques.

    We propose a GPU fine-grained load-balancing abstraction that decouples load balancing from work processing and aims to support both static and dynamic schedules with a programmable interface to implement new load-balancing schedules. Prior to our work, the only way to unleash the GPU's potential on irregular problems has been to workload-balance through application-specific, tightly coupled load-balancing techniques.

    With our open-source framework for load-balancing, we hope to improve programmers' productivity when developing irregular-parallel algorithms on the GPU, and also improve the overall performance characteristics for such applications by allowing a quick path to experimentation with a variety of existing load-balancing techniques. Consequently, we also hope that by separating the concerns of load-balancing from work processing within our abstraction, managing and extending existing code to future architectures becomes easier.
\end{abstract}

%% 2012 ACM Computing Classification System (CSS) concepts
%% Generate at 'http://dl.acm.org/ccs/ccs.cfm'.
%% End of generated code

%% Keywords
%% comma separated list

%% \maketitle
%% Note: \maketitle command must come after title commands, author
%% commands, abstract environment, Computing Classification System
%% environment and commands, and keywords command.
\maketitle

\section{Introduction}
\label{sec:introduction}

Graphical Processing Units (GPUs) excel at and are often designed for regular fine-grained parallel problems, such as General Matrix Multiplication (GEMM)\@. In regular problems like GEMM, neighboring threads have similar or identical workloads and often achieve nearly 100\% of peak GPU theoretical performance. What is much more challenging is an application with ample \emph{fine-grained} parallelism but \emph{irregular} parallelism. In such applications, neighboring threads running in a lockstep fashion will have different workloads---perhaps different amounts of work---making an efficient implementation on a highly parallel machine like a GPU a significant challenge.
% \jowens{The second half of the sentence does not follow from the first half without the important fact that nearby GPU threads run in lockstep. Work that in there? Important in an intro where you don't assume the reader is an expert.}

Consider Sparse-Matrix Vector Multiplication (SpMV), with a sparse matrix $\textbf{A}$ and a dense vector $x$ as inputs. SpMV computes the output vector  $y = \textbf{A}x$ and is an example of irregular fine-grained parallelism. Unlike in GEMM, the sparse matrix in SpMV can contain irregularity within the rows of the matrix: the rows of the matrix can have different numbers of non-zero entries. A simple mapping of one row to each GPU thread can expose this irregularity, where neighboring threads may be assigned different amounts of non-zeros to process, causing threads within the same warp\footnote{A CUDA warp is a collection of 32~threads that execute instructions in lockstep. Threads in a warp are divergent-free, and run in a Single Instruction Multiple Thread (SIMT) fashion.} to wait on threads with large amounts of non-zeros. The imbalance created due to this irregularity---specifically, when the work is not equally distributed among the parallel actors, and consequently, some actors are idle while others do more work---is defined as the load-imbalance problem.

% \jowens{You need to be much more precise about what the ``load imbalance problem'' is. Consider that this paper might become \emph{the} reference on GPU load imbalance. It is vital that you define it precisely. Separate this paragraph into two: one that defines load imbalance in a complete but succinct way and one that discusses how it is currently implemented on GPUs. It is OK if the GPU-specific treatment is not complete here as long as it provides a good-enough mental model for a reader to understand and it forward-references the part of the paper where you do talk about GPU specifics, presumably in the background section.}
Current implementations solve this load-imbalance problem on GPUs using application-specific load-balancing techniques that aim to evenly distribute the work such that each thread gets the same number of work items to achieve maximum performance (for instance, Merrill and Garland's load-balanced SpMV implementation~\cite{Merrill:2016:MPS}). These load-balancing techniques are often tightly coupled with the application itself.
% \jowens{The second half of the preceding sentence belongs in the ``what is load imbalance'' paragraph, not the ``how load imbalance is currently addressed'' paragraph.}
The load-balancing components within these implementations are both complex and often collectively the most significant contributor to the performance of an application. Our work here generalizes today's application-specific load-balancing algorithms into a clean, modular, powerful abstraction that can be applied to many complex irregular workloads.

In the process of building our abstraction, we identified common load-balancing approaches currently deployed with\-in sparse, irregular applications on GPUs: application-specific frameworks such as GraphIt~\cite{Brahmakshatriya:2021:CGA}, Gunrock~\cite{Wang:2017:GGG}, and Graph\-BLAST~\cite{Yang:2021:GAH}; techniques from low-level CUDA libraries such as ModernGPU~\cite{Baxter:2013:MPA} and CUB~\cite{NVIDIA:2023:CUB}; and other hand-coded implementations of load-balancing algorithms within applications such as SpMV/SpMM~\cite{Merrill:2016:MPS,Davidson:2014:WPG,Gale:2020:SGK}, triangle counting~\cite{Fox:2019:ISI,Green:2018:LRB}, and breadth-first search~\cite{Merrill:2012:SGG,Busato:2015:BAE}.
% \jowens{You would do well to have a broader set of citations here that go beyond our coauthors. (Your reference list doesn't count against your page budget at PPoPP\@!) You might just instead reference what's surely going to be a table in this paper of the techniques you're implemented and/or the work where those techniques were introduced/used.} \jowens{I still think adding more cites here is good. merge-path? log binning? GraphIt? The larger diversity of cites we show here, the more a reviewer knows that we know what we're talking about. Cites take almost no space.}
We show that with a simple, intuitive, powerful abstraction, these load-balancing schedules can be extended to support irregular workloads that are more general than the specific problem for which they were designed. We demonstrate this by using sparse-linear-algebra-based load balancing for data-centric graph traversal kernels.

Writing high-performance load-balancing code is complex, in large part because this code must perform many roles. Among other tasks, it must ingest data from a specific data structure, perform user-defined computation on that data, and schedule that computation in a load-balanced way. The key insight in our abstraction is to separate the concerns between workload mapping (the load-balance task) and work execution (the user-defined computation), where we \emph{map} sparse formats (such as Compressed Sparse Row (CSR)) to simple abstraction components called work \textbf{atoms}, \textbf{tiles}, and \textbf{sets}. These fundamental components are expressed as composable \cpp{} ranges and range-based for loops, and are used to build load-balancing schedules. Programmers can then use these APIs to build load-balanced, high-performance applications and primitives. Expressed in this way, we can reconstruct existing application-dependent load-balancing techniques that address irregularity to be more \emph{general}, \emph{portable}, and \emph{programmable}.
% Our abstraction allows programmers to focus on writing the actual computation rather than the underlying load-balancing algorithms, therefore, introducing code portability, reducing complexity, and improving overall performance characteristics of irregular-parallel workloads through quick experimentation.
The contributions of our work are as follows:

\begin{enumerate}
    \item We present a novel abstraction for irregular-parallel workloads on GPUs. Our abstraction at a high level allows programmers to develop sparse, irregular-parallel algorithms with minimal code while delivering high performance.
    \item We design and implement a set of intuitive APIs, available in our open-source GPU load-balancing framework, built on the proposed abstraction using CUDA-\cpp{} ranges and range-based for loops.
    \item We show the ease of implementing new load-balancing schedules by implementing a novel cooperative groups-based load-balancing schedule, described in Section~\ref{sec:load-balancing-schedules}, which is a generalization of previous thread-, warp-, and block-level load-balancing schedules~\cite{Yang:2018:DPF}.
    \item We provide state-of-the-art SpMV performance as a benchmark with a geomean of speedup of 2.7$\times$ for the SuiteSparse Matrix Collection~\cite{Davis:2011:TUO} over cuSparse's state-of-the-art implementation using simple heuristics and 3~GPU load-balancing schedules.
\end{enumerate}

\section{Design Goals}
\label{sec:design-goals}
Our programming model focuses on the broad category of fine-grained nested data parallelism. Load-balancing task-level parallelism requires a different approach and is beyond the scope of this work. This section highlights the design goals of our load-balancing abstraction:

\paragraph{Achieve high performance.} First and foremost, the goal of our work is to achieve the high performance of existing-load balancing algorithms for irregular applications. Our abstraction cannot come at the cost of significant overhead or performance degradation. We measure our success in achieving high performance by comparing the performance of our abstraction against the performance of existing hardwired implementations.

\paragraph{A composable and programmable interface.} Importan\-tly, we do not want to restrict the user to a library interface that takes control of the larger system. Programmers strongly prefer to adopt new software components that fit into their control structures rather than require them to adopt a new control structure. We want to allow the users to (1) maintain control of GPU kernel boundaries (kernel launches), (2) be able to add new load-balancing algorithms, and (3) compose new load-balanced primitives from existing load-balancing APIs. We measure the programmability of our work by comparing the Lines of Code (LOC) of our abstraction against existing implementations and show composability by implementing a new load-balancing algorithm in terms of our existing APIs.

\paragraph{Extensible to new applications.} We aim to decouple and extend application-specific load-balancing techniques to new irregular-parallel domains. Our abstraction seeks to promote the reuse of existing load-balancing techniques for new applications. We use SpMV as a benchmark application implemented using three different load-balancing techniques, some of which were previously used to implement parallel graph analytics kernels~\cite{Davidson:2014:WPG,Brahmakshatriya:2021:CGA,Wang:2017:GGG,Busato:2015:BAE}.

\paragraph{Facilitate the exploration of optimizations.} A key goal of our abstraction is to facilitate the exploration of optimizations for a given application by switching the underlying load-balancing algorithms used to balance the work. We want to encourage our users to experiment with heuristics and new load-balancing techniques to discover what works best for their application needs. We measure the success of this goal by optimizing SpMV's performance response for a large corpus of sparse matrices across several different load-balancing techniques.

\subsection*{Non-Goals}
In addition to the above design goals, we also define our non-goals:

\paragraph{Targeting other parallel architectures.} Although we believe the lessons learned should apply to other parallel architectures, we explicitly target NVIDIA's CUDA architecture and programming model~\cite{NVIDIA:2022:CUDA}. Many components of our abstraction leverage CUDA's compute hierarchy of threads, warps and blocks mapped onto the physical streaming multiprocessors, the oversubscription model of assigning more work than the number of processors to fully saturate the underlying hardware, and CUDA's Cooperative Groups programming model~\cite{Harris:2017:CG}, described in Section~\ref{sec:load-balancing-schedules}, to achieve high performance.

% \jowens{Suggest you add sentences about (a) what's in CUDA \emph{that you use} that is not in similar environments and/or (b) the necessary features that would be required to implement this elsewhere. We'd like reviewers to know that we've thought about this, and especially we want reviewers who are on a different platform to know what it would take (and what's missing) to implement it on their own platforms.}

\paragraph{Multi-GPU support.} This work focuses on load-imbalance issues for a single GPU and does not consider multi-GPU single-node or multi-node systems, although these are interesting directions for future work. %\serban{Let's consider citing Yuxin's Atos and Vyse work? (see Atos~\cite{Chen:2022:AAT} and Vyse~\cite{Chen:2022:SIP} for some potential approaches)}

\section{Our Load-Balancing Abstraction}
\label{sec:abstraction}

\begin{figure*}
    \centering
    \includegraphics[width=\textwidth]{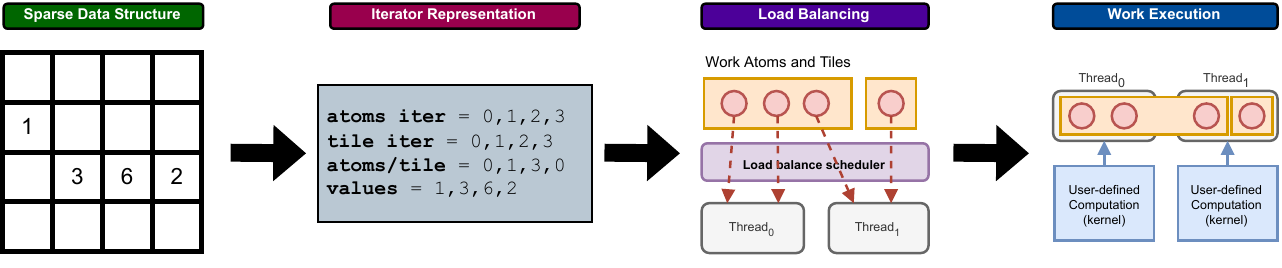}
    \caption{Load balancing as a simple pipeline of the three key concepts of our abstraction: (1)~sparse data structures represented as iterators, (2)~load-balancing algorithm that partitions the work onto threads, and (3)~user-defined computation consuming the balanced work and executing on each thread.
    % \jowens{Good figure, although there's almost no load-balancing to be done on the example. I mean, if we assign one row per thread, we get the same efficiency. Add another element to the right of 6? Also make the ``load balance scheduler'' arrows in the third figure a different color than the work atoms arrows.} \mosama{Should be better.}
    \label{fig:load-balance-abstraction}}
\end{figure*}

The key insight behind our GPU load balancing abstraction is the \emph{separation of concerns} between the mapping of the work items to processing units and work execution.
We divide our abstraction into three key concepts (illustrated in Figure~\ref{fig:load-balance-abstraction}), each of which describes a different aspect of an implementation: (1)~defining the work;
% and topology\jowens{first time the word ``topology'' has been used, not clear what that is, you do well to go through the whole doc and make sure the terminology is consistent across the entire paper, this is not a time to pull out the thesaurus};
(2)~defining the workload balance across GPU threads, warps or blocks; and (3)~defining the work execution and computation per thread on the balanced work.
This separation allows us to cleanly divide the work between an application developer and a load-balanced-library developer and facilitates the exploration of optimizations by mixing different load-balancing techniques and sparse-irregular algorithms. Sidebar~\ref{sidebar:cub-sidebar} presents a practical example of the motivation for our load balancing abstraction.
% \jowens{Maybe a sentence here to indicate \emph{why} this separation is a good idea (divide work between application developer and load-balanced-library developer, mix-and-match LB techniques, etc.). And you can specifically tie why-this-is-a-good-idea to the design goals.}

% \jowens{Any figures in this section need to be pictures, not code. The goal is for readers to understand what we are doing in this section, but not at all how we express it in C++.}

\subsection{Input from Sparse Data Structures}
\label{sec:work-domain}

% \jowens{I rewrote this paragraph, hope it's right.} \mosama{This is good!}
We begin with our input data expressed in some form of sparse data structure. Examples of such data structures include, but are not limited to, Compressed Sparse Row (CSR) and Coordinate (COO) formats. The goal of the first stage of our abstraction is to map the input data format to a common data framework and vocabulary that is the input to the next stage. This vocabulary has three simple components that together express the input data:

\begin{enumerate}
\item A \textbf{work atom}, a single unit of work that is to be scheduled onto the processors (for example, a non-zero element of a sparse matrix). We assume that all work atoms have an equal cost during execution. \jowens{Are they parallel? Tiles are parallel; are atoms within a tile parallel?} \mosama{Depends on the application; this is where most of the fine-grained parallelism is available.}
\item A \textbf{work tile}, a logical entity represented as a set of work atoms (for example, a row of a sparse matrix). Work tiles may have different costs during execution. As we highlighted in the introduction, work is most \emph{logically} parallelized over work tiles but is often most \emph{efficiently} parallelized over work atoms, and mapping between work tiles and work atoms may be expensive and complex.
\item A \textbf{tile set}, a set of work tiles that together comprise the entire working problem (for example, a sparse matrix). In our abstraction, the tiles within a tile set must be independent (and thus can run in parallel across multiple processors).
\end{enumerate}

% \jowens{This paragraph returns to abstraction. Keep it here.}
This mapping between sparse formats and atoms/tiles/tile sets is defined by the user. Though we have not implemented all of them, we believe our mapping abstraction here is flexible enough to express a wide variety of existing sparse data formats in the literature~\cite{Filippone:2017:SMM} in such a way that they are suitable for load balancing in our abstraction's next stage. As well, we have already included several common sparse formats (CSR, CSC, COO) in our load-balancing library implementation so that users can simply select and use them without having to implement them. Given a mapping to atoms/tiles/tile sets, we can next implement a load-balancing algorithm that can parallelize over work atoms or tiles transparently from the computation's perspective.

\begin{sidebar_box}[label=sidebar:cub-sidebar]{A practical example of the existing, predominant approach to load-balancing sparse-irregular workloads.}
    Consider an SpMV implementation on the GPU provided in the open-source CUDA CUB library~\cite{NVIDIA:2023:CUB}. CUB implements and maintains the SpMV algorithm presented in the paper by Merrill and Garland~\cite{Merrill:2016:MPS}. Merge-based SpMV, explained in detail in Section~\ref{sec:merge-path-lb}, is a CSR-based, perfectly load-balanced SpMV, where each thread gets an even share of work, and the amount of work is defined by the total number of matrix rows and the total number of non-zeros, summed. In the reference, this highly efficient, state-of-the-art implementation took 1,100~lines of code (LoC) (or 503~LoC of kernel code) across 3~files (not including a 4th file required for a segmented fixup step of an additional 234 LoC). In contrast, the actual computation of SpMV within this reference implementation is expressed within a \emph{single} for-loop and 4--5 LoC\@! This disparity between the LoC required to map the work items to processing units in a load-balanced way and the LoC required to express the desired computation is the key motivation behind our work. Additionally, the CUB implementation is specifically dedicated to the SpMV algorithm and would require a significant rewrite to apply it to other algorithms, even within the same computing domain. One such example of this exact rewrite is by Yang et al., who extend merge-path load balancing from SpMV to a Sparse-Matrix Dense-Matrix Multiplication (SpMM) implementation~\cite{Yang:2018:DPF}. The load-balancing algorithm in both works is the same but applied to different computations, which motivates the need for reuse.
\end{sidebar_box}

\subsection{Defining Load Balancing}
\label{sec:load-balancing-domain}

%sdp (1)~defining the work;
% (2)~defining the workload balance across GPU threads, warps or blocks; and (3)~defining the work execution and computation per thread on the balanced work.
% This separation allows us to distinctly divide the work between an application developer and a load-balanced-library developer and facilitates the exploration of optimizations by mixing different load-balancing techniques and sparse-irregular algorithms.
% \serban{ %section 4.2
By expressing workloads through an abstraction that captures work at differing levels of granularity (i.e., tile set, atoms, and tiles), we can more easily distribute computation evenly across the GPU's available resources. Given a user-defined input tile set and associated sequences of atoms and tiles, along with a user-selected partitioning algorithm, our load-balancing stage outputs subsequences of atoms and tiles assigned to processor ids (i.e., where atoms or tiles will be processed).

The resulting assignment of subsequences to processor ids is critical to effectively balancing workloads across processing elements and is generally problem- and dataset-specific. The user must specify the necessary sequences. Ideally, an \emph{oracle} would take these sequences and select the most optimal subsequences for every processing element. Finding such an oracle is an open problem and thus we provide the next best thing: the ability for users to choose and experiment from a set of predefined schedules and the ability to implement their own schedules. In general, load-balancing algorithm designers must balance between the cost of scheduling and the benefits from better scheduling. A schedule could be as straightforward as assigning processing elements to tiles with arbitrary numbers of atoms (e.g., rows with an arbitrary number of non-zeros in a sparse matrix) to something more complicated/expensive that takes on a more holistic approach to work (e.g., considering work across multiple rows with a varying number of non-zeros in a sparse matrix).

\subsection{Defining Work Execution}
\label{sec:work-execution}

The final component of our load-balancing abstraction expresses the irregular-parallel computation itself. The previous stage inputs load-imbalanced work and load-balances it; this stage then consumes that load-balanced work by performing computation on it. The scope of what computation can be expressed is extensive, and is only limited by how the load-balanced work, represented as sequences,
% \jowens{can you pick a different term than ``iterable sequences'' here?}
can be consumed within a CUDA kernel. Since the framework does not assume control of the kernel, anything you can write in a CUDA kernel will also work in our framework. For instance, programmers can express a mathematical operation performed on each atom or each tile of the work, or build cooperative algorithms that not only consume the work assigned to each thread but also combine the results with neighboring threads to implement more complex algorithms such as parallel reduce or scan. Practical examples that we have implemented in our framework (see Section~\ref{sec:work-execution-representation} and~\ref{sec:application-space-and-optimization-choices}) using this abstraction include, but are not limited to, sparse-linear algebra kernels, such as Sparse-Matrix and Sparse-Tensor contractions, and data-centric parallel graph algorithms, such as Single-Source Shortest Path (SSSP) and Breadth-First Search (BFS) built on a neighborhood traversal kernel.

We expect typical users of our library will \emph{only} write their own code for \emph{this} stage of the abstraction and use standard data structures and load-balancing schedules that are already part of our library. However, those users can also implement custom data formats and load-balancing schedules.

% \begin{figure*}
%     \centering
%     \includegraphics[width=\textwidth,trim={0 2.5cm 0 5.9cm},clip]{figs/abstraction.png}
%     \caption{} \label{fig:abstraction}
% \end{figure*}

\section{High-Level Framework Implementation}
\label{sec:framework}

Our GPU load-balancing framework implements the abstraction described in Section~\ref{sec:abstraction} using \cpp{}17 and CUDA\@. In our system, programmers use CUDA/\cpp{} to develop irregular-parallel algorithms and implement new load-balancing schedules. Per our design goals of composable APIs, extensibility, and reuse,
this and the following section introduce the implementation details of our API, and how it is used to develop new applications that promote the reuse of high-performance load-balancing techniques available within the framework. We also explore a new load-balancing method (Section~\ref{sec:load-balancing-schedules}) built on CUDA's Cooperative Groups model. Furthermore, we identify how our work can be used to facilitate the exploration of optimizations for a given application such as SpMV\@.

\subsection{Implementing Sparse Data Structures}
\label{sec:sparse-data-structure-representation}

Our framework translates sparse data structures (e.g., COO, CSR, CSC) into work atoms, work tiles, and tile sets (Section~\ref{sec:work-domain}) using simple \cpp{} iterators. \cpp{} iterators are objects that point to some element in a range of elements and enable iteration through the elements of that range using a set of operators. For example, a \lstinline{counting_iterator} is an iterator that represents a pointer into a range of sequential values~\cite{CPPREFERNCE}. Our framework requires the user to define three important iterators using \cpp: (1)~an iterator over all work atoms; (2)~an iterator over the work tiles; and (3)~an iterator over the number of atoms in each work tile. (Our library already supports several common sparse data structures.)
% \jowens{What is the difference between (1) and (3)?} \mosama{clarified.}
Using these iterators, the load-balancing schedule can then determine and distribute load-balanced work across the underlying hardware. Listing~\ref{lst:csr-tile-set} shows how our abstraction expresses the commonly used CSR format as a tile set within our framework. %ed iterable sequences for each thread to consume, which can be achieved using \cpp{} ranges.

\begin{listing}
  \caption{Compressed-Sparse Row (CSR) format expressed within our framework using C++17.
  % Using C++17, we define a counting iterator used to iterate over the atoms and tiles, and a transform iterator used to iterate over the atoms in each tile set.
  The CSR format describes a matrix using three arrays: (1)~column indices of nonzero values; (2)~the extent of rows (row offsets); and (3)~the nonzero values. Since the CSR data structure does not contain arrays that point to indices of atoms and tiles (nonzeros and rows), in the listing above we define atom and tile iterators as simple counting iterators from 0 to the total number of nonzeros ($\textit{nnz}$) and from 0 to the total rows in the matrix ($\textit{rows}$), respectively (Lines~2--3). The iterator over the atoms-per-work-tile is expressed using a transform iterator, which computes the expression within a provided function for each tile id. For CSR, this is simply the row offset of the current tile subtracted from the offset of the next tile (Lines~5--11).
  % \jowens{I just copied the text into the caption, where I think it belongs; the detail of iterator formation does not contribute to the implementation body but instead ought to be associated with the code in the caption, I think. Anyway, this probably needs some reorg.}
  }

  % \jowens{I don't like this caption. ``as a tile set'' doesn't mean anything to me. This should be in implementation, not the abstraction. Honestly, you can probably move the body text into the caption because what the reader needs to know is that we've expressed three iterators here and what they do. Also it looks like magic how \texttt{atoms\_per\_tile} works, I don't actually know how it works. Finally you might note this is Actual C++ As Of Year 20XX because old people like me look at this and think, wow, that doesn't look like C++ at all. Once you get to implementation, you might note how modern C++ and its iterators really make some complex compute tasks simple to express. In this case, the transform appears to do a lot of heavy lifting (though I don't know how it works).}
  % \mosama{Improved in text and caption.}
  \label{lst:csr-tile-set}
  \begin{minted}[
      fontsize=\small,
      obeytabs=true,
      tabsize=4,
      linenos=true,
      frame=lines,
      numbersep=-1pt]{c++}
  // Simple iterators for atoms and tiles.
  counting_iterator<int> atoms_iter(0, nnz);
  counting_iterator<int> tile_iter(0, rows);
  // Iterator over the atoms within tile i.
  auto atoms_per_tile = make_transform_iterator(
    tile_iter,
    [tile_iter, row_offsets]
    __host__ __device__(const int& i) {
      return (row_offsets[tile_iter[i + 1]] -
              row_offsets[tile_iter[i]]);
  });
  \end{minted}
\end{listing}

\subsection{Implementing Load-Balancing Schedules}
\label{sec:load-balancing-representation}

Perhaps the most straightforward schedule is scheduling each work tile onto one GPU thread. This approach is common in the literature and practice~\cite{Merrill:2012:SGG,Davidson:2014:WPG,Steinberger:2017:GHL,Baxter:2013:MPA,Yang:2018:DPF}; although this strategy is ineffective in the presence of significant load imbalance across tiles, we use it here as an example to illustrate how load balancing is defined within our framework.

% \jowens{Starting at ``Each'' and going up to but not including the last sentence, this should probably go into the caption.}
The inputs are the three iterators from the last stage plus an atom and tile count. The load-balance algorithm developer, then, implements \lstinline{tiles()} and \lstinline{atoms()} procedure calls, which return the \cpp{} range of tiles and atoms to be processed by the current thread, effectively creating a map between assigned processor ids and segments of the workload. Listing~\ref{lst:thread-mapped} shows a complete example of the thread-mapped schedule. Although a simple algorithm, it can deliver high performance for well-balanced workloads with coarse-grained parallelism (a small number of atoms per tile), such as multiplying a sparse vector by a dense vector. Furthermore, our abstraction is not limited to only simple scheduling algorithms, as Section~\ref{sec:load-balancing-schedules} provides examples of more complex load-balancing algorithms.
% \jowens{What you \emph{need} to describe in the body text of this subsection is generically what this stage looks like. The inputs, for instance, are the three iterators from the last stage plus an atom and tile count. You have to define tiles and atoms procedure calls. What do they do, generically? You want the reader to know ``this is what I have to specify in this stage of the abstraction'' and that's not clear from what's above.}
% \jowens{Now that I'm reading this, it's way too many words to describe something really simple from the abstraction point of view. This is probably not the right example. I think you can take a sentence or two to explain, basically, that a mapping of $n$ work tiles to $n$ threads is simple/trivial. Maybe next you try to explain something a little more complicated. But the important part is then to give an indication of, genetically, what kind of schedules the abstraction can express. What is the scope of the abstraction? Do this without reference to iterators. Do this without reference to particular scheduling algorithms, though after you give this explanation, you can say ``and we can do merge-path and TWC etc.'' and cite them. What needs to be said at this point in the paper is ``what is the scope of schedules we can express''.}

% \jowens{I think I would stop this paragraph here (``provides examples of numerous more complex scheduling algorithms.'') and put these cites in Section~\ref{sec:load-balancing-schedules}.}

\begin{listing}
  \caption{A thread-mapped load-balancing algorithm expressed as C++ ranges, incorporating the atoms and tiles defined as iterators from Listing~\ref{lst:csr-tile-set}.
  Each tile is mapped to a thread, where the thread id corresponds to the index of the tile in the tile set. All atoms within a tile are sequentially processed by the thread. After a tile is processed, a thread is mapped to the next tile, obtained by striding the index by the grid size of the kernel.
  % \jowens{What is a ``range''?} \jowens{This needs some explanation about what is happening here; a couple of sentences per procedure is probably good? The comments are helpful, but they say what the code should accomplish, and maybe the caption should say what the implementation actually does (e.g., round-robin distribution of tiles across all threads). Probably you want to move the description within the Section~\ref{sec:load-balancing-representation} body to here. Finally, this incorporates the iterators you defined in Listing~\ref{lst:csr-tile-set}, yes? Say so!} \jowens{Could you make it more clear in naming that this is a thread\_mapped schedule? Maybe \texttt{typedef thread\_mapped\_schedule\_t schedule\_t}? Then declare it as and use the more specific name?}
  % \jowens{This is implementation, not abstraction. Also it needs some explanation.}
  }
  \label{lst:thread-mapped}
  \begin{minted}[
      fontsize=\small,
      obeytabs=true,
      tabsize=4,
      linenos=true,
      frame=lines,
      numbersep=-1pt]{c++}
  class schedule_t {
    // Construct a thread-mapped schedule.
    __host__ __device__
    schedule_t(atoms_it_t atoms_it,
      tiles_it_t tiles_it,
      atoms_it_t atoms_per_tile_it,
      size_t num_atoms, size_t num_tiles) :
      m_atoms_it(atoms_it), m_tiles_it(tiles_it),
      m_atoms_per_tile_it(atoms_per_tile_it),
      m_num_atoms(num_atoms),
      m_num_tiles(num_tiles) {}
    // Range of tiles to process in "this" thread.
    // Stride by grid dimension.
    __host__ __device__ auto tiles() {
      auto begin = m_tiles_it(blockDim.x * blockIdx.x
                 + threadIdx.x);
      auto end = m_tiles_it(m_num_tiles);
      return range(begin, end)
             .step(gridDim.x * blockDim.x);
    }
    // Range of atoms to process in "this" thread.
    __host__ __device__ auto atoms(
        const std::size_t& tile) {
      auto begin = m_atoms_per_tile_it[tile];
      auto end = m_atoms_per_tile_it[tile + 1];
      return range(begin, end).step(1);
    }
  };
  using schedule_t = thread_mapped_schedule_t;
  \end{minted}
\end{listing}

\subsection{Implementing Work Execution}
\label{sec:work-execution-representation}

% \jowens{I think this benefits from a little more discussion. I would start with a potential design decision of ``let the framework own the kernel'' and then explain what the pros/cons of that decision are, and why you turned to an alternate design. Explaining design decisions is really important for systems papers!}

Our framework is designed to explicitly let the user \emph{own} the kernel launch boundary. Owning a CUDA kernel boundary means that the user is responsible for maintaining and configuring launch parameters and implementing the CUDA kernel used to define the application. Although this design decision comes at a cost of convenience and simplicity, it offers significant flexibility in what users can express through our abstraction. This design decision is motivated by the following reasons. (1)~Users are not required to add a complex dependency to their existing workflow/libraries, therefore making code maintenance simpler and more scalable as they do not have to rely on our framework to incorporate new CUDA constructs and features. (2)~Users are free to express anything and everything CUDA allows within their kernels while consuming our load-balanced \cpp{} ranges. This allows for versatility in what can be expressed, as the users can now specify multiple load-balanced work domains, range-based for loops, and even fusing multiple computations to build more complex algorithms within a single kernel. (3) Higher-level APIs can be used to build simpler higher-level abstractions that \emph{do} own the kernel boundary and provide simpler APIs at the cost of flexibility.

% \jowens{The above paragraph motivates the design decision. This is good. However, you never say anything about what this stage actually looks like, which is kind of the point. You only offer the below example. What are the inputs/outputs of this stage? What would a typical implementation look like (e.g., doubly-nested loop over tiles and atoms)? You can probably pull text from the below example.}

As an input to this stage, users consume the load-balanced \cpp{} ranges to implement their computation. This can be done in multiple ways, but one of the most common patterns is a nested range-based for loop that loops over all the assigned tiles and atoms ranges. Listing~\ref{lst:spmv} shows a simple example of a CUDA kernel that implements the SpMV algorithm the using CSR format and thread-mapped load-balancing algorithm described in Listings~\ref{lst:csr-tile-set} and~\ref{lst:thread-mapped}. In this example, the outer \texttt{for} loop within each thread iterates over the assigned rows of the sparse matrix (tiles), and the inner loop sequentially processes the assigned nonzeros (atoms) within each row. In Section~\ref{sec:application-space-and-optimization-choices} we implement and discuss more complex kernels and computations.

\begin{listing}
  \caption{Sparse-Vector Matrix Multiplication (SpMV) implemented within our load-balancing abstraction using range-based nested \texttt{for} loops. The sparse matrix is represented using a CSR-based format, where $x$ is the dense input vector and $y$ is the dense output vector ($y = Ax$).  Lines~9--12 use the load-balancing schedule implemented in Listing~\ref{lst:thread-mapped} and the iterators defined in Listing~\ref{lst:csr-tile-set} to construct the load-balanced work to be processed. Lines~14 and~17 show the \texttt{for} loops within each thread, which iterate over the assigned rows of the sparse matrix and sequentially process the assigned atoms within each row. Line~18 shows the actual computation performed on each work atom (nonzero), and Line~19 writes the result to the dense output vector $y$.
  }
  \label{lst:spmv}
  \begin{minted}[
      fontsize=\small,
      obeytabs=true,
      tabsize=4,
      linenos=true,
      frame=lines,
      numbersep=-1pt]{c++}
  // Implements load-balanced SpMV kernel.
  __global__ void spmv(const size_t rows,
    const size_t cols, const size_t nnz,
    const int* offsets, const int* indices,
    const float* values, const float* x,
    float* y) {
    // Configure load-balancing.
    // Input: iterators defined for CSR format.
    schedule_t config(
        atoms_iter, tile_iter,
        atoms_per_tile_it,
        nnz, rows);
    // Consume rows using a range-based for loop.
    for (auto row : config.tiles()) {
      type_t sum = 0;
      // Consume atoms using a range-based for loop.
      for (auto nz : config.atoms(row))
        sum += values[nz] * x[indices[nz]];
      y[row] = sum;
    }
  }
  // Launches SpMV kernel.
  constexpr size_t blocks = 256;
  size_t grid = (rows + blocks - 1) / blocks;
  spmv<<<grid, blocks>>>(rows, cols, nnz,
    offsets, indices, values, x, y);
  \end{minted}
\end{listing}

%%% Local Variables:
%%% mode: latex
%%% TeX-master: t
%%% End:

\section{Implementation Details}
\label{sec:implementation-details}

% \mosama{@jowens---I want the next three subsections to be a part of a new section; what should that be called? Its API details, optimizations and expressing more complex things. Previous section was let's understand the framework with simpler examples.} \jowens{You could do ``High-Level Framework Implementation'' and ``Implementation Details''?}

\subsection{Flexible, Composable CUDA-enabled Ranges}
\label{sec:composable-api-ranges}

The composability of load-balanced primitives and applications using our API is
a conscious design choice within our framework supported through the use of CUDA-enabled \cpp{} ranges. Our framework does not \emph{own} the kernel boundary (kernel launch), which forces our APIs to be focused and contained within the kernels. This allows programmers to build and maintain their own kernels while still benefiting from our framework's load-balancing capabilities. This is largely implemented using device-wide \cpp{} functions and classes tagged with CUDA's \lstinline{__device__} keyword.\footnote{A method decorated with the \lstinline{__device__} keyword allows the CUDA compiler to generate a device-callable entry point. This allows the code to be called from within kernels~\cite{NVIDIA:2022:CUDA}.} We implemented and expose several different types of specialized ranges that were particularly useful in implementing load-balanced schedules:

\begin{itemize}
    \item \lstinline{step_range}: A range that iterates from \lstinline{begin} to \lstinline{end} in steps of \lstinline{step}. Useful for defining load balancing schedules that require a custom stepping range or process a constant number of work items per thread (which can be defined using \lstinline{step}).
    \item \lstinline{infinite_range}: A range that iterates from \lstinline{begin} to infinity. Useful for defining load balancing schedules in persistent kernel mode~\cite{Zhang:2022:PKI}, where the kernel persistently runs until all work is consumed or an algorithm has converged.
    \item \lstinline{grid_stride_range}: A specialized case of step range that iterates from \lstinline{begin} to \lstinline{end} in steps of \lstinline{step} using the CUDA kernel's grid size. Also supports \lstinline{block} and \lstinline{warp} stride variants that iterate in steps of the block or warp size, respectively.
\end{itemize}

\subsection{Implementing Non-Trivial Load-Balancing}
\label{sec:load-balancing-schedules}

As we describe in Section~\ref{sec:composable-api-ranges}, we can decouple and express existing load-balancing techniques as a set of \cpp{} ranges.
% \jowens{The word ``range'' doesn't even appear in Section~\ref{sec:load-balancing-representation}. If this previous sentence is what you meant to get across in this section, I missed the point. I like the point. But Section~\ref{sec:load-balancing-representation} needs to make that crystal clear.}
To illustrate the potential of this abstraction, we begin by decoupling and expressing a state-of-the-art load-balancing algorithm known as merge-path~\cite{Green:2012:GMP} previously used for balancing CSR-based SpMV and SpMM~\cite{Merrill:2016:MPS,Yang:2018:DPF}, and implement three additional load balancing algorithms (warp-, block- and group-mapped), all of which are available in our library for programmers to use. Our new group-mapped algorithm is a tile-per-group-based schedule, where a group is defined as a collection of threads of any arbitrary size (not limited to a warp or block size). Our group-mapped schedule is a generalization of the tile-per-thread, -warp or -block schedules~\cite{Merrill:2012:SGG,Brahmakshatriya:2021:CGA} using CUDA's Cooperative Groups programming model~\cite{Harris:2017:CG}.

\subsubsection{Merge-path load balancing}
\label{sec:merge-path-lb}

In the language of a sparse matrix, merge-path assumes that each non-zero in the matrix and each new row in the matrix are an equivalent amount of work, then evenly divides $\text{nnzs} + \text{rows}$ work across the set of worker threads. Each thread then performs a 2-D binary search within the nonzero indices and row offsets of a CSR matrix to find the starting position of the row and nonzero it needs to process. Threads then sequentially process the rows and nonzeros from the starting position until they reach the end of their assigned work~\cite{Merrill:2016:MPS}.

We implement this algorithm as a load-balancing schedule in our abstraction by expressing it in two steps:
(1)~\textbf{Setup:} The initialization step of the \cpp{} schedule class computes the number of work units per thread, conducts a binary search as described above, and stores the starting position of each tile and atom in a thread-local variable.
(2)~\textbf{Ranges:} The second step of the algorithm builds the ranges for each thread to process as ``complete'' tiles and ``partial'' tiles~\cite{Merrill:2016:MPS}.
% \jowens{Why are the tiles partial? I'd think the nomenclature should target the threads, not the tiles. If this is Merrill/Garland's terminology, I guess use it, but cite it.} \jowens{I rewrote up to END below, please check for accuracy.}
If a thread's atom range lies entirely within one tile, it is ``complete'', and is processed in a simple nested loop. If a thread's range crosses a tile boundary, the thread processes its work in a separate nested loop.
\jowens{I'm sensitive to space, but showing the listing of this implementation would help the reader a lot. At least give LoC for this schedule.}

Because we decouple the load-balancing method (Section~\ref{sec:load-balancing-representation}, and above) from work execution (Section~\ref{sec:work-execution-representation}), we can use this merge-path implementation to implement not only SpMV but also any other algorithm whose work can be divided into tiles and atoms, e.g., a graph neighborhood-traversal algorithm used to implement breadth-first search~\cite{Wang:2017:GGG}. Just as importantly, the merge-path schedule is now no longer limited to a CSR-based sparse format. Supporting other formats only requires building the necessary slightly more complex iterators that are able to count atoms per tile (the computation that the CSR implementation achieves with the row offsets array in Listing~\ref{lst:csr-tile-set}). %Once translated to our \emph{atoms}, \emph{tiles} and \emph{tile set} abstraction elements, other formats can directly leverage our existing merge-path implementation. \jowens{END}

\subsubsection{Warp- and block-level load balancing}
The goal of a warp- or block-level load-balancing schedule is to assign an equal share of tiles to each warp or block, which are then sequentially processed. The work atoms within each tile will be processed in parallel by the available threads within a warp or a block. Each thread strides by the size of the warp or block to process a new work atom until the end of work is reached.

The imbalance across different processing units is left for the hardware scheduler to handle. This scheduler depends on the oversubscription model of CUDA, where the programmer can launch a larger number of warps or blocks than the GPU can physically schedule at any given time. As the processing units finish processing their work, new ones are scheduled from the oversubscribed pool~\cite{Merrill:2012:SGG,Brahmakshatriya:2021:CGA}.

\subsubsection{Group-level load balancing}
Group-level load balancing generalizes warp- and block-level schedules. Instead of requiring that group sizes are the size of a warp or block, as above, this method leverages CUDA's Cooperative Groups (CG) programming model~\cite{Harris:2017:CG} to allow programmer-specified dynamically sized groups of arbitrary size. Within these groups, the CG model permits detailed control of the group's synchronization behaviors as well as simple parallel group-level collectives such as reduce or scan. We leverage this powerful tool to implement a generalized group-level load balancing schedule, effectively giving us the warp- and block-level schedules above for free when the group size equals that of a warp or a block.

Our schedule assigns work tiles to a group, and each group looks at its equal share of tiles and computes the number of atoms for each tile and stores it in a scratchpad memory (CUDA's \emph{shared memory}). The group then performs a parallel prefix-sum, a widely used parallel algorithm that inputs an array and produces a new array where the element at any position is a sum of all previous elements~\cite{Blelloch:1990:PSA}. We use this prefix-sum array for two purposes: (1)~the last element of a prefix-sum array indicates the aggregated number of work atoms that a group has to process, and~(2) the position of each sum in the prefix-sum array corresponds to the work tile to which those atoms belong. The setup phase of the schedule builds the prefix-sum arrays per group in the scratchpad memory, and the ranged-loop of the schedule returns the atom to process in each thread. The corresponding tile, if needed, is obtained by a simple \lstinline{get_tile(atom_id)} operation, which executes a binary search within the prefix-sum array to find the tile corresponding to the atom being processed.

Relying on the CG model for this load-balancing schedule has a unique advantage of configuring the group size (effectively software constructs that directly map onto the hardware) per the shape of the problem and the underlying hardware architecture. For example, targeting GPUs where the warp size is not 32~threads (AMD's GPU architecture supports a warp size of 64~\cite{AMD:2022:HIP}) is now possible with a simple compile-time constant, or configuring the group size to perfectly align with the structure of the problem.

\subsection{Application Space}
\label{sec:application-space-and-optimization-choices}

Our work definition (Section~\ref{sec:work-domain}), composable APIs (Section~\ref{sec:composable-api-ranges}), and multiple sophisticated, high-performance load-balancing schedules (Section~\ref{sec:load-balancing-schedules}) together provide for a versatile and extensible framework with plenty of room for application-specific optimizations. In Listing~\ref{lst:spmv} we already demonstrated how to implement the SpMV algorithm using our framework. A simple and natural extension is to implement Sparse-Matrix Matrix Multiplication (SpMM)\@. Listing~\ref{lst:spmm} shows the minor change necessary, which adds another loop over the columns of the $\textbf{B}$ matrix around the existing code from Listing~\ref{lst:spmv} to implement SpMM\@. This implementation could also be extended to support Gustavson's General Sparse Matrix-Matrix Multiplication (SpGEMM), using two kernels and an allocation stage; the first kernel would compute the size of the output rows used to allocate the memory for the output sparse matrix and the second kernel would perform the multiply-accumulation.

\begin{listing}
    \caption{A simple loop wrapped around SpMV introduced in Listing~\ref{lst:spmv} allows us to represent the slightly more complex SpMM load-balanced computation.
    % \jowens{Putting new code in texttt-italics would be awesome.} \mosama{Idk how to do that with escapeinside a minted code.}
    }
    \label{lst:spmm}
  \begin{minted}[fontsize=\small,
      obeytabs=true,
      tabsize=4,
      linenos=true,
      frame=lines,
      escapeinside=||,
      numbersep=-1pt]{c++}
  // ... Inside the CUDA kernel.
  // Loop over all the assigned rows.
  for (auto row : config.tiles()) {
    // Loop over all the columns of Matrix B.
    for (auto col : range(size_t(0), B.cols)
      .stride(size_t(1))) { /// < New Loop
      float sum = 0;
      // Loop over all the assigned nonzeros.
      for (auto nz : config.atoms(row))
          sum += values[nz] * B(nz, col);
      // Output the sum to Matrix-C.
      C(row, col) = sum;
    }
  }
  \end{minted}
\end{listing}

Beyond sparse linear algebra, we can use our framework to address applications in other domains. Listing~\ref{lst:sssp} implements the graph primitive Single-Source Shortest Path (SSSP) using our group-level load-balancing schedule. SSSP's performance on GPUs is largely gated by good load balancing~\cite{Wang:2017:GGG,Brahmakshatriya:2021:CGA}, but if the programmer chooses a load-balancing schedule from our library, the details of load balancing are completely hidden. Moreover, the same schedules that were used in one application domain (e.g., sparse linear algebra) are easily reusable in this different application domain.

\begin{listing}
  \caption{The parallel single-source shortest path (SSSP) graph primitive expressed using our load-balanced schedule. \jowens{Note in this code where you select group-level load balancing.}}
  \label{lst:sssp}
  \begin{minted}[
      fontsize=\small,
      obeytabs=true,
      tabsize=4,
      linenos=true,
      frame=lines,
      numbersep=-1pt]{c++}
  // ... Inside the CUDA kernel.
  // Loop over all the assigned edges to process.
  for (auto edge : config.atoms()) {
    auto source = config.get_tile(edge);
    // G is the graph data structure
    auto neighbor = G.get_neighbor(source, edge);
    auto weight = G.get_edge_weight(edge);
    float source_dist = dist[source];
    float neighbor_dist = source_dist + weight;
    // Check if the destination node has been
    // claimed as someone's child.
    float recover_distance =
      atomicMin(&(dist[neighbor]), neighbor_dist);
    // Add the neighbor to the frontier.
    if (neighbor_dist < recover_distance)
      out_frontier[neighbor] = true;
  }

  // ... Outside the CUDA kernel.
  // Loop until the frontier is empty.
  \end{minted}
\end{listing}

% \mosama{Optimization choices WIP\ldots}

%%% Local Variables:
%%% mode: latex
%%% TeX-master: t
%%% End:

\section{Evaluation}
\label{sec:evaluation}

We aim to show that our framework, built on our load balancing abstraction, enables both high performance and better programmability for sparse-irregular problems. Our evaluation below uses our SpMV implementation as a benchmark against state-of-the-art implementations provided within NVIDIA's (open-source) CUB library and production (closed-source) cuSparse library. We considered (and implemented) several additional applications for evaluation, including SSSP, BFS, and SpMM\@. We found they led to similar high-level conclusions. Thus our evaluation here focuses on SpMV\@. Our test corpus consists of approximately the \emph{entire} SuiteSparse Matrix Collection~\cite{Davis:2011:TUO} with a broad scope of sparse matrices from many different high-performance computing domains. We ran all experiments on a Ubuntu 20.04 LTS-based workstation with an NVIDIA Tesla V100 GPU and CUDA 11.7.

\subsection{Performance Overhead}
\label{sec:performance-overhead}

\begin{figure}
    \centering
    \includegraphics[width=\columnwidth]{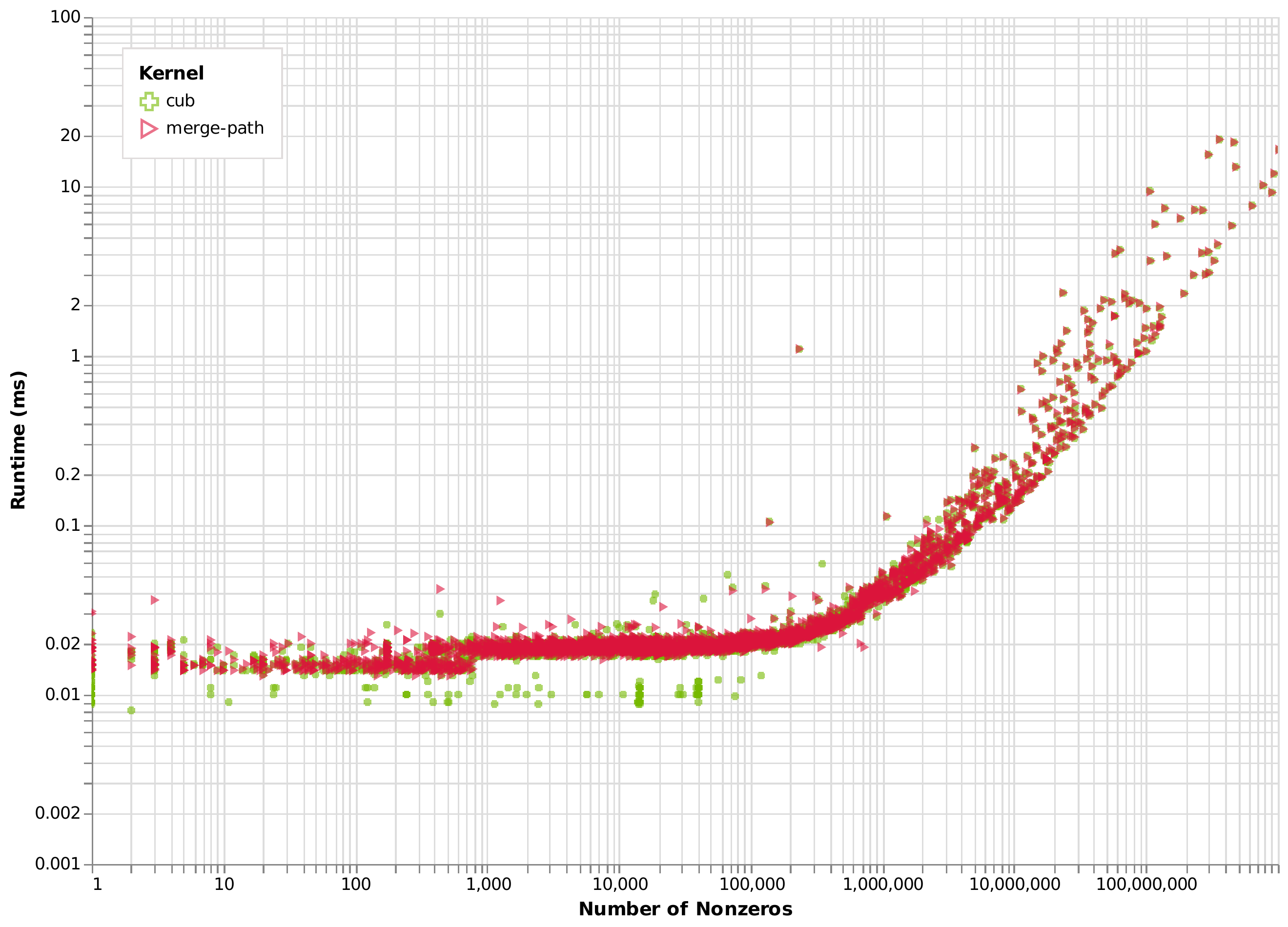}
    \caption{SpMV runtime comparison: our merge-path SpMV implementation vs.\ CUB across all SuiteSparse datasets. Our runtimes almost perfectly match CUB's for all datasets. The small number of datasets where CUB is faster is due to a simple heuristic that CUB uses for single-column sparse matrices (i.e., a sparse vector).}
    \label{fig:performance-overhead}
\end{figure}

Our first and foremost goal is to ensure that the elements within our abstraction do not add any additional performance overhead to the existing load balancing techniques and algorithms developed using them. To verify this, we compare the runtime performance of our SpMV implementation using the merge-path schedule to the implementation provided by NVIDIA's CUB library~\cite{NVIDIA:2023:CUB} (also used for Merrill and Garland's merge-path SpMV paper~\cite{Merrill:2016:MPS}) on the SuiteSparse collection. As previously mentioned, and in contrast to our design, CUB contains a hardwired implementation of the merge-path scheduling algorithms and does not decouple workload balancing from the actual SpMV computation. CUB's approach is not reusable for any other irregular parallel problem without significant changes to the implementation.

Figure~\ref{fig:performance-overhead} plots the number of nonzeros (i.e., the total work) vs.\ runtime for our work vs.\ CUB's implementation. Our implementation has minimal performance overhead when using our abstraction: a geomean slowdown of 2.5\% vs.\ CUB, with 92\% of datasets achieving at least 90\% of CUB's performance. Figure~\ref{fig:performance-overhead} shows our implementation almost perfectly matches CUB for all datasets, except for some datasets with fewer than 100,000 nonzeros. Upon further investigation, we identify that CUB uses a simple heuristic to launch a thread-mapped SpMV kernel where the number of columns of a given input matrix equals 1 (i.e., a sparse vector). Unlike our more general implementation, CUB's simple (but specialized) thread-mapped SpMV kernel has no load-balancing overhead for a perfectly balanced workload such as SpVV computation.

\subsection{Improved Performance Response}
\label{sec:improved-performance-response}

\begin{figure*}
    \centering
    \includegraphics[width=\textwidth]{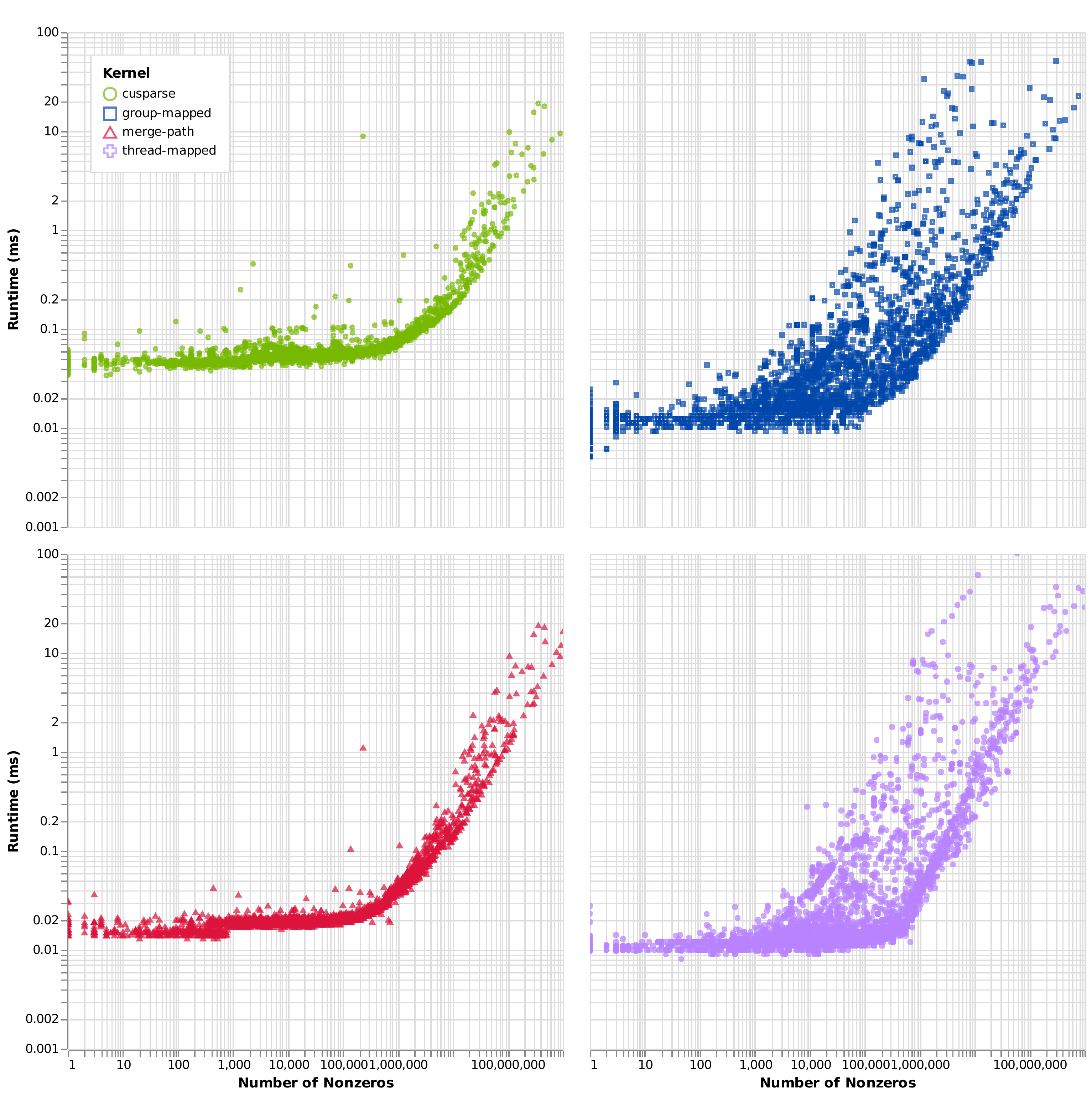}
    \caption{Complete performance landscape of SpMV across all SuiteSparse datasets using 3~load balancing schedules vs.\ NVIDIA's cuSparse library. This performance comparison highlights the impact of different approaches to load-balancing SpMV for a given dataset and number of nonzero entries within each dataset. Later in Figure~\ref{fig:spmv-speedup} we use this insight to select the fastest schedule for an improved overall performance. Additionally, our 3~different SpMV implementations are made possible with very little code change.}
    \label{fig:performance-response}
\end{figure*}

We also compare our work to NVIDIA's vendor library for sparse computations, cuSparse. Figure~\ref{fig:performance-response} shows the performance response of our SpMV implementation using each of our scheduling algorithms individually vs.\ cuSparse's state-of-the-art implementation. Switching between any of our implementations requires very little code change; in the case of merge-path and thread-mapped, we need only update a single \cpp{} enum (identifier) to select the desired load-balancing schedule.

\begin{figure}
    \centering
    \includegraphics[width=\columnwidth]{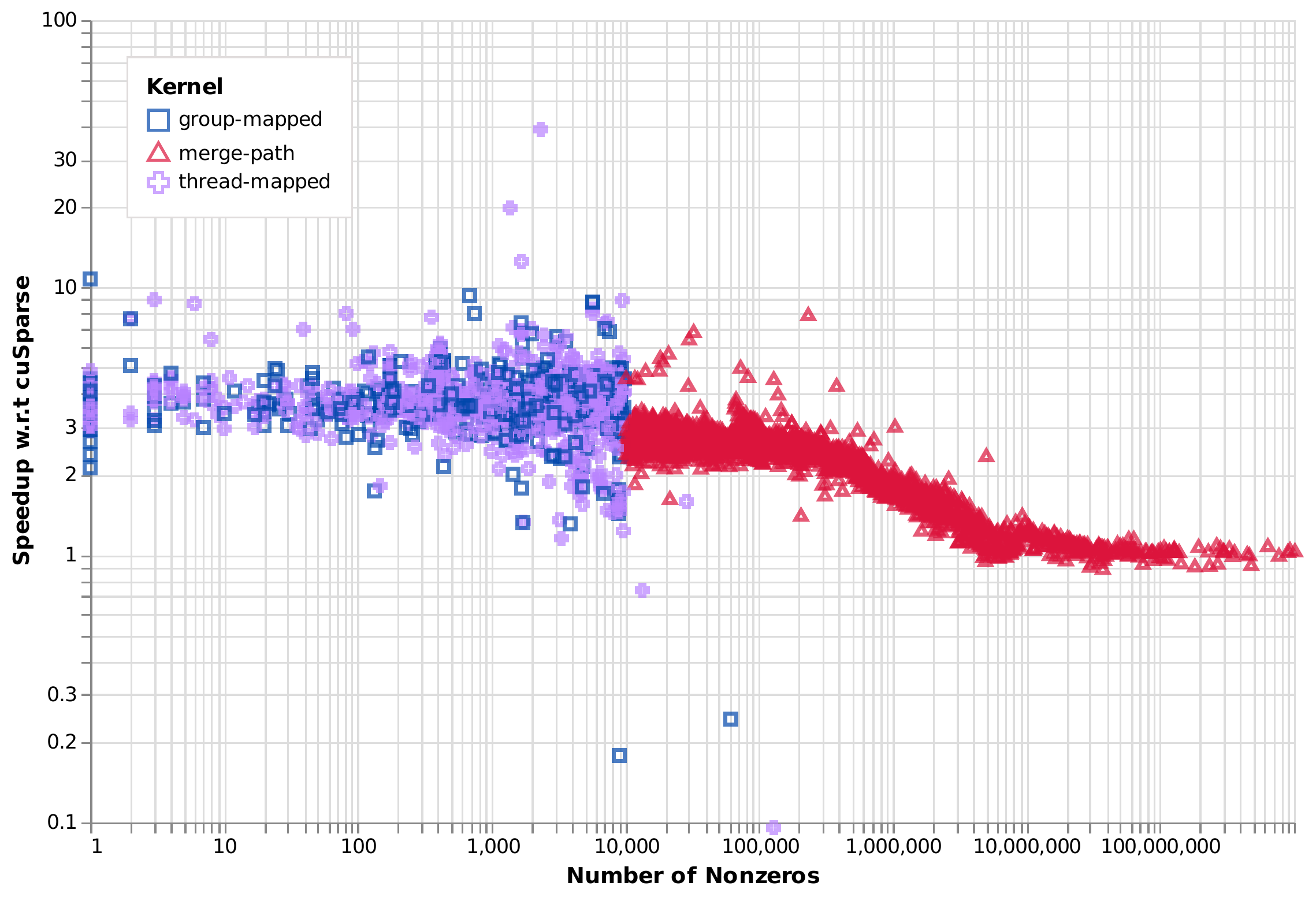}
    \caption{Speedup of our framework's SpMV vs.\ cuSparse's SpMV across SuiteSparse using a heuristic (Section~\ref{sec:improved-performance-response}) to choose the appropriate load-balancing schedule.}
    \label{fig:spmv-speedup}
  \end{figure}

  We then combine our scheduling algorithms into one implementation for SpMV (Figure~\ref{fig:spmv-speedup}), demonstrating noticeable performance improvements over cuSparse. This is primarily possible due to our ability to quickly experiment with different heuristic schemes with a variety of available load-balancing schedules. Here, we use merge-path unless either the number of rows or columns are less than the threshold $\alpha$ and the nonzeros of a given matrix are less than threshold $\beta$ (we choose $\alpha = 500$ and $\beta = 10000$ for SuiteSparse). In this case, we use thread-mapped or group-mapped load balancing instead of merge-path. Our system shows a peak performance speedup of 39$\times$ and a geomean performance speedup of 2.7$\times$ vs.\ cuSparse.

Our framework not only allows programmers to express computations efficiently and simply (i.e., without worrying about the load-balancing algorithms), but also quickly optimize a given application using a range of scheduling algorithms, both with minor code changes.

\subsection{Lines of Code (LOC)}
\label{sec:loc}

We are able to achieve these performance gains with minimal code complexity. Table~\ref{tab:spmv_loc} shows lines of code (LOC) for our framework when compared to the state-of-the-art open-source implementation of merge-path and thread-mapped within NVIDIA's CUB library. We deliver the same performance results as highlighted in the previous sections with 14$\times$ and 1$\times$ fewer lines of code for merge-path and thread-mapped scheduling algorithms, respectively. Using our merge-path implementation only requires $\sim$15 additional LoC to the trivial thread-mapped schedule.

Furthermore, we extend the same SpMV computation to our novel \emph{group-mapped} load balancing schedule (that can also be specialized to perform \emph{block-} and \emph{warp-mapped} load balancing) within the same 30 LoC\@.

\begin{table}
  \centering
  \begin{small}
    \begin{tabular}{@{}lcc@{}}
      \toprule
      Load Balancing Algorithm & NVIDIA/CUB & Our Work \\ \midrule
      Merge-Path & 503 & 36 \\
      Thread-Mapped & 22 & 21 \\
      Group-Mapped & N/A & 30 \\
      Warp-Mapped & N/A & 30 (free) \\
      Block-Mapped & N/A & 30 (free) \\
      \bottomrule
    \end{tabular}
    \end{small}
  \setlength{\belowcaptionskip}{0pt}
  \caption{Lines of code (LoC) comparison for NVIDIA's CUB library versus our work for SpMV application implemented using merge-path, thread-mapped and group-mapped (warp- and block-mapped use the \emph{exact} same code for group-mapped) load balancing algorithms. We report only non-commented lines of code, formatted using the \lstinline{clang-format} tool with the Chromium style guide~\cite{Google:Chromium}, that contributes to the kernel implementation.}
  \label{tab:spmv_loc}
\end{table}

\section{Related Work}
\label{sec:related-work}

Load balancing is the key to achieving high performance on GPUs for sparse, irregular parallel problems. Several high-performance computing applications deploy sophisticated load balancing algorithms on the GPUs. For instance,  high-performance sparse-matrix vector multiplication (SpMV) leverages merge-path~\cite{Merrill:2016:MPS} (discussed in detail in this paper) or a nonzero splitting algorithm, which partitions the number of non-zeros in a sparse-matrix evenly across the number of threads~\cite{Baxter:2013:MPA,Dalton:2015:OSM,Steinberger:2017:GHL}. Sparse-matrix matrix multiplication (SpMM) and sparse matricized tensor times Khatri-Rao product (SpMTTKRP) use binning and bundling algorithms~\cite{Yang:2018:DPF,Gale:2020:SGK,Nisa:2019:LSM}, which attempt to bin like-length work together such that they are processed together.

While some applications actively perform work to load-balance a given input, others store the input in more efficient, already-load-balanced/-partitioned formats. These include the F-COO format (a variant of coordinate format) used for SpMTTKRP and Sparse-Tensor Tensor Multiplication (SpTTM), where each thread gets the same number of nonzeros to process~\cite{Liu:2017:UOA}.

Many of the above GPU load-balancing algorithms, along with other novel techniques, were first described in the graph analytics domain. Davidson et al.\ and Merrill and Garland were the first to present Warp, Block-level and Thread-Warp-CTA dynamic load balancing techniques for Single-Source Shortest Path (SSSP) and Breadth-First Search (BFS) respectively~\cite{Davidson:2014:WPG,Merrill:2012:SGG}. Logarithmic Radix Binning (LRB) is a particularly effective technique for binning work based on a logarithmic work estimate, used for the Triangle Counting graph algorithm and more~\cite{Green:2018:LRB,Fox:2019:ISI}. Gunrock, GraphIT, and GraphBLAST are graph analytics libraries that implement several different graph algorithms such as BFS, SSSP, PageRank, Graph Coloring, and more, built on these previously mentioned load-balancing techniques~\cite{Wang:2017:GGG,Brahmakshatriya:2021:CGA,Yang:2021:GAH}. Although many of these are effective load balancing techniques with high-performance implementations, they all tightly couple workload scheduling with the application itself. Our framework is designed to separate these two concerns, allowing the application to be independent of the load-balancing algorithm, and therefore be expressed simply. Our approach also allows these previously proposed techniques to be implemented within our framework, and be used for applications beyond those originally targeted.

Relatively few GPU works target generalized load balancing for irregular workloads. Most of these are focused on providing a singular, dynamic load-balancing solution centered on task parallelism, often using a GPU queue-based data structure. Cederman and Tsigas proposed a task-based approach to load balancing an octree partitioning workload using lock-free and lock-based methods~\cite{Cederman:2008:ODL}. Two Tzeng works provide task-management frameworks that implement load balancing of tasks using a single monolithic task queue and distributed queues with task stealing and donation~\cite{Tzeng:2010:TMF,Tzeng:2012:AGT}.
% \jowens{I think the previous sentence conflates two Tzeng works~\cite{Tzeng:2010:TMF,Tzeng:2012:AGT}; the 2010 work was distributed queues, the 2012 work was a single queue.}
CUIRRE, a framework for load balancing and characterizing irregular applications on GPUs, also uses a task-pool approach~\cite{Zhang:2014:CAO}, and more recently, Atos, a task-parallel GPU dynamic scheduling framework, targets asynchronous algorithms~\cite{Chen:2022:AAT}. \jowens{This paragraph is long and could be simplified dramatically if space required.} All of these works deploy either a centralized or a distributed queue-like data structure on the GPUs, each making design decisions on how the queue is to be partitioned and updated. Except for the most recent Atos work, most earlier works focus on a coarse-grained parallelism approach of effectively distributing tasks to the GPU\@. Our work takes advantage of more modern GPU architectures, which are more effectively utilized by a fine-grained parallelism approach (parallelizing over work atoms instead of work tiles). Unlike our abstraction, these aforementioned works also rely on a singular load-balancing solution, whereas our abstraction flexibly adapts to many different load-balancing techniques, static and dynamic, and allows for new schedules to be implemented within our framework.

% % \begin{verbatim}
% - Dense-Regular Parallel Problems to
% Sparse-Irregular Problems
% - Fine-grained versus Coarse-grained parallelism
% - Significance of Load Balancing

% - NVIDIA's CUDA Platform
% - Modern \cpp{}
% % \end{verbatim}

% \subsection{Related Work}
% % \begin{verbatim}
% - Taxonomy of Load Balancing
% - Summary of Load Balancing Work + CPU reference
% - Domain-specific works
%   - GraphIT
%   - A performance, power, and energy efficiency
%   analysis of load balancing techniques for GPUs
%   - Configuring Graph Traversal Applications for
%   GPUs: Analysis of Implementation Strategies and
%   their Correlation with Graph Characteristics
%   - Improving Scheduling for Irregular
%   Applications with Logarithmic Radix Binning
% % \end{verbatim}

\section{Conclusion}
\label{sec:conclusion}

In this paper, we present a programming model for GPU load balancing for sparse irregular parallel problems. Our model is built on the idea of separation of concerns between workload mapping and work execution. In the future, we are interested in expanding our model to a multi-GPU environment, and implementing load-balancing schedules that span across the GPU boundary covering multiple devices and nodes for massive parallel problems.
Our current work focuses solely on load balancing, but we also identify locality to be another key factor for high performance. We are interested in identifying an orthogonal model that builds an abstraction for caching and locality into our existing load-balancing framework.

%% Acknowledgments
\begin{acks}                            %% acks environment is optional
    This material is based upon work supported by \grantsponsor{}{Defense Advanced Research Projects Agency (DARPA)}{} under Contract No.~\grantnum{}{HR0011-18-3-0007} and the \grantsponsor{}{National Science Foundation}{} under Contract No.~\grantnum{}{OAC-1740333}. Any opinions, findings and conclusions or recommendations expressed in this material are those of the author(s) and do not necessarily reflect the views of the U.S.\ Government. Distribution Statement ``A'' (Approved for Public Release, Distribution Unlimited). We would like to acknowledge Michael Garland and Duane Merrill from NVIDIA for their guidance on the framework. We would also like to acknowledge Toluwanimi Odemuyiwa, Jonathan Wapman, Matthew Drescher and Muhammad Awad for research discussions and feedback on the work. We also acknowledge the support of AMD, Inc. (Jalal Mahmud and AMD Research) in the form of travel funding, which enables us to attend the conference to present this work.
\end{acks}

%% Bibliography
\bibliography{bib/all.bib,missing.bib}

%% Appendix
\newpage
\appendix
\section{Artifact Description}
\label{sec:appendix}

We provide the source code of our load-balancing framework called \lstinline{loops} and our testing harness for evaluating the results provided within this paper.

\subsection{Requirements}

\begin{enumerate}
    \item \textbf{Operating System} Ubuntu 18.04, 20.04, Windows.
    \item \textbf{Hardware} NVIDIA GPU (Volta microarchitecture or newer).
    \item \textbf{Software} CUDA 11.7 or above and cmake 3.20.1 or newer.
    \item \textbf{Compilation} NVCC (comes with CUDA), g++ and gcc, msvc with support for \cpp{}14 standard.
    \item \textbf{Output} Comma-separated values (CSV) files that are used to generate the graphs in Section~\ref{sec:evaluation}.
    \item \textbf{Disk space} 886~GB to store the entire SuiteSparse Matrix Collection~\cite{Davis:2011:TUO} compressed and uncompressed. Can be reduced significantly by running the tests on only a subset of the dataset.
    \item \textbf{Code License} Apache 2.0.
    % \item **CUDA Architecture:** SM 70 or above (see [GPUs supported](https://en.wikipedia.org/wiki/CUDA#GPUs\_supported)), this is specified using cmake's command: \lstinline{-DCMAKE_CUDA_ARCHITECTURES=70}.
\end{enumerate}

\subsection{How to Access}
The main repository is hosted on GitHub: \url{https://github.com/gunrock/loops}. Our framework is also available as a Zenodo archive: \url{https://doi.org/10.5281/zenodo.7465053}~\cite{Osama:APM:2022:Zenodo}. Detailed and well-formatted instructions are available within the README markdown file in the repositories, and a summary is available below.

\subsection{Getting Started}
Before building \lstinline{loops}, make sure you have the CUDA Toolkit and cmake installed on your system, and exported in \lstinline{PATH} of your system. Other external dependencies such as \lstinline{thrust}, \lstinline{cub}, etc.\ are automatically fetched using cmake.

\begin{lstlisting}[language=bash,basicstyle=\ttfamily\small]
cd loops
mkdir build && cd build
cmake -DCMAKE_CUDA_ARCHITECTURES=70 ..
make -j$(nproc)
\end{lstlisting}

\subsubsection{Sanity Check}
Run the following command in the cmake's \lstinline{build} directory:
\begin{lstlisting}[language=bash,basicstyle=\ttfamily\small]
bin/loops.spmv.merge_path \
 -m ../datasets/chesapeake/chesapeake.mtx \
 --validate -v
# Expected Output
# Elapsed (ms):   0.063328
# Matrix:         chesapeake.mtx
# Dimensions:     39 x 39 (340)
# Errors:         0
\end{lstlisting}

\subsection{Reproducing Results}
We provide the following instructions to regenerate the results presented in this paper.

\begin{enumerate}
    \item In the run script, update \lstinline{DATASET_DIR} to point to the path of all the downloaded datasets (set to the path of the directory containing the \lstinline{MM} directory; inside \lstinline{MM} are subdirectories with \lstinline{.mtx} files): \lstinline{scripts/run.sh}.
    \begin{itemize}
      \item You may change the path to \lstinline{DATASET_FILES_NAME} containing the list of all the datasets (default points to \lstinline{suitesparse.txt} file in the \lstinline{datasets} directory).
    \end{itemize}
    \item Fire up the complete run using \lstinline{run.sh} found in the scripts directory, \lstinline{cd scripts && ./run.sh}. Note one complete run can take up to 3 days (the run goes over the entire SuiteSparse matrix collection dataset four times with four different algorithms; the main bottleneck is loading files from disk).
    \begin{itemize}
    \item Warning: Some runs on the matrices are expected to fail as they are not in proper MatrixMarket Format although labeled as \lstinline{.mtx}. These matrices and the ones that do not fit on the GPU will result in runtime exceptions or type overflow and can be safely ignored.
    \end{itemize}
    \item To run \lstinline{N} number of datasets, simply adjust the stop condition here (default set to 10): \lstinline{run.sh#L22}, or remove this if-condition entirely to run on all available \lstinline{.mtx} files: \lstinline{run.sh#L22-L26}.
\end{enumerate}

Additionally, we provide pre-generated results (in the form of CSV files) to create the plots from Section~\ref{sec:evaluation} without needing to run all the experiments. These pre-generated results are available under the \lstinline{docs} directory of the repository.

\subsection{Expected Output and Plots}

The expected output from the above runs are csv files in the same directory as the \lstinline{run.sh}. These can replace the existing csv files within \lstinline{docs/data}, and a python jupyter notebook can be used to evaluate the results. The python notebook includes instructions on generating plots. See a sample output of one of the csv files below:
\begin{lstlisting}[language=bash,basicstyle=\ttfamily\small]
kernel,dataset,rows,cols,nnzs,elapsed
merge-path,144,144649,144649,2148786,0.07202
merge-path,08blocks,300,300,592,0.0170898
merge-path,1138_bus,1138,1138,4054,0.0200195
\end{lstlisting}

\end{document}